\documentstyle[12pt]{article}
\begin{document}

{\bf Loutsenko and Roubtsov Reply:} 
The model proposed  in \cite{LR} can be considered as a 
model  describing  the anomalous 
transport of nonrelativistic bosons
in a medium with the linear dispersion law
at zero temperature.

The main idea of  Letter
\cite{LR}  is  that  the effective
two-particle
interaction between bosons switches from being repulsive to
the attractive one 
if the  velocity of collective propagation of the
bosons in the medium exceeds some critical speed.
This effect occurs  due to different covariance of the dynamic equations 
that describe the nonrelativistic subsystem
of bosons and the linearly dispersive medium. As a result, 
the boson subsystem ``collapses'' into a soliton wave packet above this
threshold.

Note that the effect of  such an anomalous
propagation is the quantum one, since Bose-Einstein condensation 
(more exactly, occurrence of the exciton-phonon Bose condensate)
is essential in our consideration.

The theory developed in \cite{LR} was applied to interpret the experiments
on excitonic propagation in semiconducting crystals (Cu$_{2}$0) at nonzero
temperature \cite{AM}.
In this 
series of experiments, a crystal was irradiated 
by  laser pulses. 
At low intensities of the laser beam (i.e. at low concentration of
the excitons), 
the system revealed  a typical {\it diffusive} behavior. 
Once the intensity of
the beam exceeded some value, the majority of particles moved together in a
sharp {\it solitonic} packet with
the ballistic velocity exceeding some critical speed. This coincides with 
the main result of our theory.

Although this theory yields a qualitative description of the experiments and
a reasonable value for the critical velocity, the estimate of the width of the 
condensate at zero temperature is
in a strong disagreement with experimental data \cite{AM} for the total
exciton-phonon packet (at finite temperature).
Indeed, we obtain \cite{LR} 
\begin{equation}
L_{\rm ch}(N_{\rm o},\,v) \simeq 
4\,
\sqrt{ \frac{ \hbar^{2}}{ m_{\rm x}\,\vert \tilde{\nu}_{0}(v)\vert }\,
\Phi_{\rm o}^{-2}\bigl(N_{\rm o},\,v\bigr)\,}
 \label{estimate1},
\end{equation}
i.e.,  $L_{\rm ch} = {\cal F}\,a_{\rm x}$, where the large factor ${\cal F}$
can vary as $10^{2} \sim 10^{4}$, 
and $a_{\rm x}$ is the exciton Bohr radius.
Thus, the duration of the  condensate can be estimated as   
$t_{\rm ch} \simeq 2 \times (10^{-11} - 10^{-9})$\,s 
(compare with 
the corresponding estimate in \cite{ST}). 
On the other hand, 
$\Delta t \simeq 5\times 10^{-7}$\,s  was obtained experimentally
\cite{AM}. 

This fact 
was pointed out by S.~G.~Tikhodeev in Comment \cite{ST}.
In our opinion, the crucial question is
whether the Bose-Einstein condensate, or, better, 
any macroscopically occupied
coherent mode,  exists inside the exciton-phonon packet at $T<T_{c}$.
If yes, one can ask, for example, 
how many excitons form the coherent core 
of the packet, 
and the value of $N_{\rm o}(T)/N_{\rm tot}$ has to be estimated at 
$T<T_{c}$. 

Indeed,  localized moving solutions for 
the exciton concentration $n(x,t)$  
can be obtained within classical models (see, e.g., \cite{ST2}), 
in which 
the Bose-Einstein condensate of excitons is absent
and the excitonic cloud is
dragged by the sound wave of a  large amplitude 
created under the action of the strong laser pulse.

Here, we list several facts
that support  the idea that the localized excitonic condensate 
has to be taken into account to interpret experimental data \cite{AM}.

\newpage

\begin{itemize}
\item 
Experiments are conducted at nonzero temperature, and  the excitonic condensate 
(if appears)
can constitute of a relatively small fraction of the total number of particles. 
We considered
the  system  bosons\,$+$\,medium at zero temperature.
Experimental data, however, 
show strong dependence of the packet length 
on the temperature
(at least, an order of magnitude  in the range of $2 - 5$\,K). 
Thus, 
no conclusion can be made until
our theory is extended to nonzero temperatures or experiments are 
conducted at much more lower temperatures.

\item 
All the results on nonlinear interaction between two packets 
\cite{AM} point
out to a  kind of coherent interaction.
This is expected within our quantum model 
of the moving exciton-phonon droplet with 
the ``Bose-nucleus''. 
It is an open question whether such a behavior can be the case within any 
classical model.

\item
If, according to experiments,  
the  phonon source is generated at a surface by a strongly absorbed
radiation, the phonon wind \cite{ST2} 
does not effect strongly the excitons with a density just below
the threshold one, and 
no appreciable effect is detected. 
On the contrary, the injection of
cold excitons distributed throughout the volume leads to the appearance of a
localized packet in the above conditions.
\end{itemize}

We believe that the  future theory needs both the exciton-phonon condensate
and  the proper incorporation of non-condensed excitons and  phonons.
It can be tested by further experiments as well. 
For example,  
one can set the
crystal geometry in such a way that the sound wave 
of the phonon wind theory can be dumped out, and only
the coherent part of the packet (if exists) will continue to propagate.

In conclusion, we note that  
the phonons play a crucial role 
in almost all current  models aimed to explain or predict  
coherent  behavior of excitons in semiconductors, see, f.ex., 
\cite{Lozovik},\cite{AIm},\cite{Ivanov}.
The fact that our simple model 
fails to predict the width of the packet correctly
does not mean our approach is wrong.
Indeed,  one has to take into account the thermal excitons 
(e.g., the weak tail that is always observed behind the soliton \cite{AM}) 
and the thermal phonons of the crystal 
to make the model with {\it condensate} more realistic. 


\vspace*{0.8cm}
I. Loutsenko, $^{1}$ and D. Roubtsov $^{2}$
\newline
$^{1}$Physics Department
\newline
Princeton University
\newline
Princeton New Jersey 08544 
 \vspace*{0.2cm}
\newline
$^{2}$D\'epartement de Physique et GCM
\newline 
Universit\'e de Montr\'eal
\newline
Montreal, Quebec, Canada H3C 3J7
\newline

\vspace*{0.3cm}

Received September 1999
\newline


\end{document}